\documentclass[pra,superscriptaddress,twocolumn,showpacs,floatfix]{revtex4}
\usepackage{graphicx}
\usepackage{epstopdf}
\usepackage{amsmath}
\usepackage{color}
\usepackage{bm}

\begin{document}
\title{Time-resolved measurement of Landau--Zener tunneling in different bases}

\author{G. Tayebirad\footnote{g.tayebirad@thphys.uni-heidelberg.de}}
\affiliation{Institut f\"ur Theoretische Physik, Universit\"at Heidelberg, Philosophenweg 19, 69120 Heidelberg, Germany}

\author{A. Zenesini}
\affiliation{Institut f\"ur Experimentalphysik, University of Innsbruck, Technikerstrasse 25, 6020 Innsbruck, Austria}

\author{D. Ciampini}
\affiliation{CNISM-Pisa,  Dipartimento di Fisica, Universit\`{a} di 
Pisa, Lgo Pontecorvo 3, 56127 Pisa, Italy}
 \affiliation{INO-CNR, Dipartimento di Fisica, Universit\`{a} di 
Pisa, Lgo Pontecorvo 3, I-56127 Pisa,Italy}

\author{R. Mannella}
\affiliation{CNISM-Pisa, Dipartimento di Fisica,  Universit\`{a} di 
Pisa, Lgo Pontecorvo 3, 56127 Pisa, Italy}

\author{O. Morsch}
 \affiliation{INO-CNR, Dipartimento di Fisica, Universit\`{a} di 
Pisa, Lgo Pontecorvo 3, I-56127 Pisa,Italy}

\author{E. Arimondo}
\affiliation{CNISM-Pisa, Dipartimento di Fisica, Universit\`{a} di 
Pisa, Lgo Pontecorvo 3, 56127 Pisa, Italy}
 \affiliation{INO-CNR, Dipartimento di Fisica, Universit\`{a} di 
Pisa, Lgo Pontecorvo 3, I-56127 Pisa,Italy}

\author{N. L\"orch}
\author{S. Wimberger}
\affiliation{Institut f\"ur Theoretische Physik, Universit\"at Heidelberg, Philosophenweg 19, 69120 Heidelberg, Germany}

\begin{abstract}
A comprehensive study of the tunneling dynamics of a Bose--Einstein condensate in a tilted periodic potential is presented. We report numerical and experimental results on time-resolved measurements of the Landau--Zener tunneling of ultracold atoms introduced by the tilt, which experimentally is realized by accelerating the lattice. The use of different protocols enables us to access the tunneling probability, numerically as well as experimentally, in two different bases, namely, the adiabatic basis and the diabatic basis. The adiabatic basis corresponds to the eigenstates of the lattice, and the diabatic one to the free-particle momentum eigenstates. Our numerical and experimental results are compared with existing two-state Landau--Zener models.

\end{abstract}


\date{\today}

\maketitle

\section{Introduction}

Quantum transport is an essential topic in solid state physics and electronic applications. Bloch oscillations, Landau--Zener (LZ) tunneling, and Wannier--Stark ladders~\cite{Landau,Zener,Stueckelberg,Majorana32,Leo92,Dahan-Peik,Wilkinson,Anderson98,Glueck02}, are fundamental quantum effects occurring in a system of electrons moving in a periodic potential and driven by an electric field. Due to complications such as impurities, lattice vibrations, and multiparticle interactions, clean observations of these effects have been difficult~\cite{Leo03}. In recent years, ultracold atoms and Bose--Einstein condensates in optical lattices have been increasingly used to simulate solid state systems and the above mentioned phenomena~\cite{Dahan-Peik,Wilkinson, niu98,Anderson98,Grynberg01,Morsch01,Roati04,Morsch06,Bloch08}.\\
\indent  Optical lattices are easy to realize in the laboratory, and their parameters can be perfectly controlled both statically and dynamically, which makes them attractive as model systems for crystal lattices. The LZ model for transitions~\cite{Landau,Zener} between two energy states at an avoided level crossing is one of the few exactly solvable examples of time-dependent quantum mechanics. LZ transitions have been investigated for Rydberg atoms~\cite{Rubbmark81}, molecular nanomagnets~\cite{Wernsdorfer99,Foldi07}, field-driven superlattices~\cite{Sibille98}, current-driven Josephson junctions~\cite{Mullen88}, Cooper-pair box qubits~\cite{Oliver-Sillanpaa}, and using light waves in coupled waveguides~\cite{Khomeriki05,Longhi05,Dreisow09}. While the asymptotic tunneling probability can be calculated accurately~\cite{Holthaus00} and has an intuitive interpretation as a statistical mean value of experimental outcomes, the concept of tunneling time and its computation are still the subject of debate even for simple systems~\cite{Mullen88,mullen_89,Berry_90,vitanov_96,vitanov_99,Schulman07}. The tunneling time is the time required for a state to evolve into an orthogonal state.\\
\indent In this paper, we present numerical as well as experimental results on the Wannier--Stark system. This system is realized with ultracold atoms, forming a Bose--Einstein condensate, in an optical lattice subjected to a static tilting force~\cite{Anderson98}. The tilt is experimentally implemented by accelerating the optical lattice~\cite{Dahan-Peik,Wilkinson,Morsch01,Morsch06,Cristiani02,JonaLasinio03,Sias07,Zenesini08,Zenesini09}. We explore the LZ tunneling between the Bloch bands of a Bose-Einstein condensate in such an accelerated lattice. The lattice depth controls the tunneling barrier, while its acceleration controls the time dependence of the Hamiltonian. At large accelerations LZ tunneling leads to significant interband transitions for the condensate \cite{Morsch06,Holthaus00}. This tunneling process is detected by measuring the atomic momentum distributions. \\
\indent Following our previous work, in which we presented time-resolved observations of LZ tunneling~\cite{Zenesini09}, in the present article we report more detailed investigations. We measure the time dependence of the tunneling probability by performing a projective quantum measurement on the eigenstates in a given basis of the Hamiltonian describing the Bose--Einstein condensate within the optical lattice.
Our measurements resolve the steplike time dependence of the occupation probability. Using different numerical as well as experimental protocols, we are able to perform our calculations and experiments both in the adiabatic basis of the lattice eigenstates and in the diabatic basis of the free-particle momentum eigenstates. We present theoretical and experimental results which clearly show that the time dependence of the transition probability exhibits a steplike structure with a finite transition time and oscillations with a finite damping time, all of them depending on the choice of the measurement basis. To our knowledge, such time-resolved measurements in different bases have not been reported for other systems before.\\
\indent  The paper is organized as follows. Section II collects the necessary theoretical background to describe the probability and transition time for the LZ transition tunneling. The limits one faces in applying this theory to the Wannier--Stark problem, and the essential theoretical and numerical tools to describe our time-resolved measurements are reported in Section III. Section IV presents numerical and experimental data. We discuss and summarize our results in Section V.

\section{Survival probability and transition time}
\subsection{LZ theory in a nutshell}

Quantum mechanical systems having two discrete energy levels are omnipresent in nature. For crossing levels, there is a possibility of a transition if the degeneracy is lifted by a coupling and the system is forced across the resulting avoided crossing by varying the parameter that determines the level separation. This phenomenon is known as a LZ tunneling. LZ theory, developed in the early 1930's in the context of atomic scattering processes and spin dynamics in time-dependent fields~\cite{Landau,Zener,Stueckelberg,Majorana32}, demonstrated that transitions are possible between two approaching levels as a control parameter is swept across the point of minimum energy separation. The phase accumulated between the incoming and outgoing passages varies with, e.g., the collision energy, giving rise to St\"uckelberg oscillations in the populations~\cite{Stueckelberg}.

In its basic form the LZ problem can be described by a simple two-state model and allows for a simple expression for the transition probability. The LZ Hamiltonian for a single crossing taking place at time $t=0$ can be written as the following $2$ by $2$ matrix
\begin{equation}
  H_{\rm LZ}=\left(
    \begin{array}{cc}
         \alpha t & \Delta E/2 \\
          \Delta E/2 &  -\alpha t\\
    \end{array}
  \right) \ .
  \label{eqno1}
\end{equation}

 \begin{figure}[htc] 
 \begin{center}
 \includegraphics[width=0.75\linewidth,angle=0]{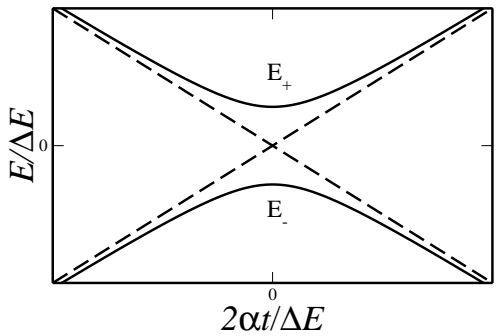}
 \caption{ \small \label{fig:1} Energy levels as a function of time. The dashed lines show the so-called diabatic levels, i.e., the energies of states in the absence of the interaction. The solid lines demonstrate the so-called adiabatic levels, i.e., the eigenstates of the system corresponding to the instantaneous Hamiltonian.}
 \end{center}
 \end{figure}
The off-diagonal term, $\Delta E/2$, is the coupling between the two states and $\alpha$ is the rate of change of the energy levels in time. The dynamics of the system can be measured in different bases, {\it diabatic} and {\it adiabatic}. The diabatic basis is the eigenbasis of the bare states of Eq.~\eqref{eqno1} when there is no off-diagonal coupling. The adiabatic basis, on the other hand, is the basis of a system with a finite $\Delta E/2$ coupling between the two states. The Hamiltonian has two adiabatic energy levels $E_{\pm}=\pm\frac{1}{2}\sqrt{\left(2\alpha t\right)^2 + \Delta E^{2}}$.\\
\indent  Assuming that the system is initially, at $t\rightarrow -\infty$, in the ground energy level $E_{\rm -}$ and if the sweeping rate is small enough, it will be exponentially likely that the system remains in its adiabatic ground state $E_{\rm -}$ at $t\rightarrow +\infty$. The limiting value of the adiabatic  LZ survival
probabilities (for $t$ going from $-\infty$ to $+\infty$) is~\cite{Holthaus00},
\begin{equation}
{P_{\rm a}(\infty)=1-\exp\left(-\frac{\pi}{\gamma}\right),
\label{eqno2}}
\end{equation}
where we introduce a dimensionless parameter, the so called adiabaticity parameter $\gamma=4\hbar\alpha/\Delta E^{2}$.
This survival probability is valid for both $E_-$ and $E_+$ initial states, and the same equation is valid for the diabatic case. A small adiabaticity parameter corresponds to a small velocity of the state displacement along the energy scale compared to $\Delta E^{2}$, such that the system follows the adiabatic trajectory of Fig.~\ref{fig:1}. Thus, there is a large coupling between the diabatic states and at the avoided crossing at $t=0$ an almost complete transition from the initial diabatic state to the final diabatic state takes place. On the other hand, for a large value of the adiabaticity parameter $\gamma$, the coupling between the two states is small and consequently the system remains in its initial state following the diabatic trajectories of Fig.~\ref{fig:1}.

\subsection{Jump times}

A careful study of the transition from an initial state to a final state can reveal the time required to complete the transition. Moreover, in the case of multiple level crossings, as in our experimental realization of ref.~\cite{Zenesini09}, it is necessary to know whether a transition has been completed before the next avoided crossing. The LZ approach may be applied when a transition between two coupled quantum states takes place in a small time interval around the avoided crossing and successive crossings are independent from each other.

\begin{figure}[htc]
 \begin{center}
 \includegraphics[width=1\linewidth,angle=0]{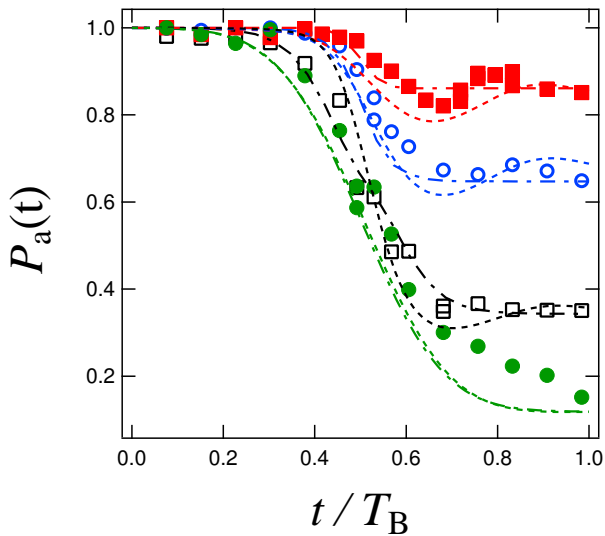}
 \caption{ \small \label{fig:adiab_exp} (Color online) Time-resolved measurements of the adiabatic survival probability for the Bose--Einstein condensate tunneling in an optical lattice at fixed dimensionless force $F_0=1.197$ and different lattice depths: $V/E_\mathrm{rec}=2.3$ (filled squares), $1.8$ (open circles), $1$ (open squares) and $0.6$ (filled circles); all parameters are introduced in detail in section IIIA. For the Bose--Einstein condensate evolution the crossing time is $t=T_{\rm Bloch}/2$, where the step of the survival probability is one half of the final value. The dashed and dot-dashed lines are the results of numerical simulations using the cut-off and adiabatic method, respectively (see section IIIC). The lattice depth for the numerical simulations was corrected by up to $\pm 15\%$ with respect to the experimentally measured values to give the best possible agreement.}
 \end{center}
\end{figure}

Analytical estimates for the LZ transition times have been derived in~\cite{vitanov_96,vitanov_99} using the two-state model of Eq.~\eqref{eqno1}. In a given basis, e.g., adiabatic or diabatic, different transition times are obtained. Vitanov~\cite{vitanov_99} calculated the time-dependent diabatic/adiabatic survival probability at finite times. The LZ transition times were derived in~\cite{vitanov_96} using some exact and approximate results for the transition probability. Fig.~\ref{fig:adiab_exp} shows a typical time dependence of the adiabatic survival probability, similar to that predicted in~\cite{vitanov_99}, that we measured for Bose--Einstein condensates in optical lattices for experimental parameters to be discussed in Section~\ref{results}. Notice that in the Bose--Einstein condensate case the crossing occurs at the time $t=T_{\rm B}/2$ defined below. The $t\to\infty $ asymptotic value is given by Eq.~\eqref{eqno2}.

The LZ jump time in a given basis can be defined as the time after which the transition probability reaches its asymptotic value. From this definition one can expect to observe a step-like structure, with a finite width, in the time-resolved tunneling probability, as in Fig.~\ref{fig:adiab_exp}. Because the step is not very sharp, it is not straightforward to define the initial and final times for the transition. It is even less obvious how to define the jump time for both small and for large coupling. Some possible choices have been used by Lim and Berry~\cite{Berry_90} and Vitanov~\cite{vitanov_96, vitanov_99}. The problem is even more complicated when the survival probability shows an oscillatory behavior on top of the step structure as seen in  Fig.~\ref{fig:adiab_exp}, which shows experimental and numerical results for a single LZ transition measured in the adiabatic basis (the numerical and experimental methods will be described in detail later in this paper). The oscillations give rise to other time scales in the system, namely an oscillation time and a damping  time of the oscillations appearing in the transition probability after the crossing. Therefore, a measurement of the tunneling time depends very much on how these times are defined and also which basis is considered.\\
 \indent  In~\cite{vitanov_99}, the jump time in the diabatic/adiabatic bases is defined as
\begin{equation}
\tau^{jump}_{\rm d/a}=\frac{P_{\rm d/a}(\infty)}{P'_{\rm d/a}(0)}.
\label{eqno3}
\end{equation}
where $P_\mathrm{d/a}$ is the transition probability between the two diabatic/adiabatic states, respectively. $P'_{\rm d/a}(0)$  denotes the time derivative of the tunneling probability evaluated at the crossing point. The diabatic jump time $\tau^{jump}_{\rm d}\approx \sqrt{2\pi\hbar/\alpha}$ is almost constant for large values of the adiabaticity parameter $\gamma$~\cite{vitanov_99}. Instead, for $\gamma\ll1$ it decreases with increasing $\gamma$, $\tau^{jump}_{\rm d}\approx 2\sqrt{\hbar(\gamma\alpha)^{-1}}$~\cite{vitanov_99}. In the adiabatic basis, when $\gamma$ is large the transition probability resembles the one of the diabatic basis with an equal jump time. For a small adiabaticity parameter, because of the oscillations appearing on top of the transition probability step structure, it is not straightforward to define the initial and the final time for the transition. Vitanov~\cite{vitanov_99} defines the initial jump time as the time $t<0$ at which the transition probability is very small (i.e., $P_{\rm a}(\tau)=\varepsilon P_{\rm a}(\infty)$, where $\varepsilon$ is a proper small number). The final time of the transition $t>0$ is defined as the time at which the non-oscillatory part of $P_{\rm a}(\tau)$ is equal to $(1+\varepsilon)P_{\rm a}(\infty)$. Using these definitions, Vitanov derived  that the transition time in the adiabatic basis depends exponentially on the adiabaticity parameter, $\tau^{jump}_{\rm a}\approx \left(4/\epsilon\right)^{1/6}\gamma^{-1/3}\mathrm{exp}\left(\pi/(6\gamma)\right)\sqrt{\hbar/\alpha}$,~\cite{vitanov_99}.

In principle, the experimental and numerical methods presented in the following could be used for a quantitative study of the tunneling time (or jump time) as function of the parameters of the system. For the purposes of the present paper, however, we concentrate on a careful analysis of the possibilities and limitations of our methods, and in particular on measurements of LZ tunneling in different bases.

\section{LZ in an optical lattice potential}
\subsection{Wannier--Stark problem and LZ limit}

We generalize the two-level LZ theory to study the temporal evolution of ultracold atoms loaded into a spatially periodic potential subjected to an additional static force in the presence of negligible atom-atom interactions, as in the experimental conditions~\cite{Zenesini09}. The dynamics of ultracold atoms in a tilted optical lattice can be described by the well-known Wannier--Stark Hamiltonian~\cite{Glueck02}
\begin{equation}
{\tilde H=-\frac{\hbar^{2}}{2M}\frac{d^2}{dx^2}+\frac{V}{2} \cos\left(2k_{\rm L} x\right)+F_{\rm LZ} x,
\label{WannierStark}}
\end{equation}
where $M$ is the atomic mass, $V$ is the depth of the optical lattice with the spatial period $d_{\rm L}=\lambda_{\rm L}/2$ determined by the laser wavelength $\lambda_{\rm L}$, $k_{\rm L}=2\pi/\lambda_{\rm L}$ is the wave number of the laser light creating the periodic potential, and $F_{\rm LZ}$ is the Stark force. The characteristic energy scale of the system is the recoil energy which is defined as $E_{\rm rec}=\pi^{2}\hbar^{2}/2 M d_{\rm L}^{2}$.

\begin{figure}[htc]
 \begin{center}
 \includegraphics[width=1\linewidth,angle=0]{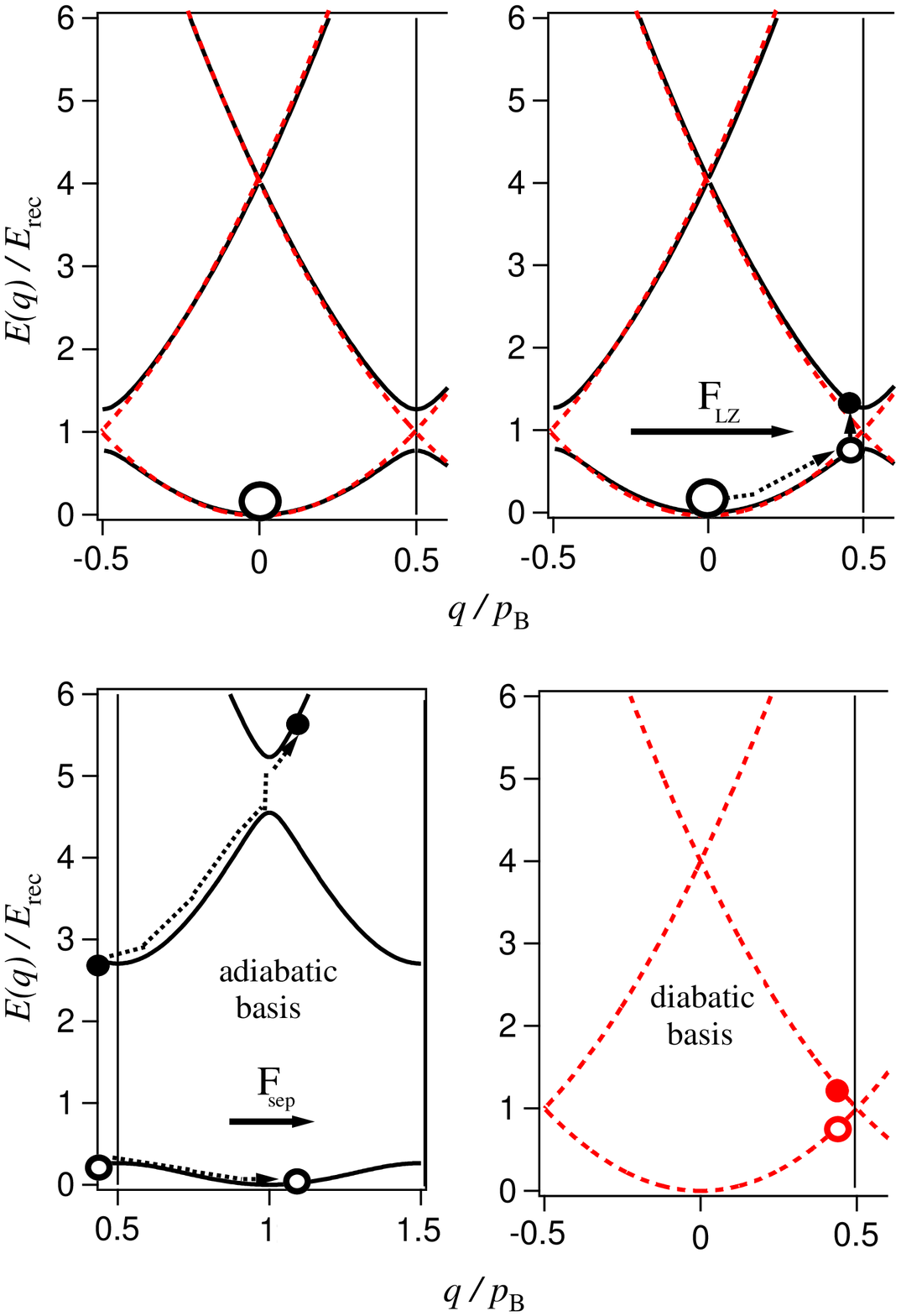}
 %{expscheme.eps}
 \caption{\small \label{fig:expscheme} (Color online) Band structure in the optical lattice potential and experimental protocols for measuring LZ dynamics in the adiabatic and diabatic bases. After the initial loading into the lattice and acceleration for a time $t_\mathrm{LZ}$ (top), measurements of the instantaneous populations in the two states are performed (bottom) as explained in the text. In the top figures, the adiabatic (solid lines) and diabatic energies (dashed lines) for an optical lattice of depth $V_\mathrm{0}=1$ are shown.}
\label{energyscheme}
 \end{center}
 \end{figure}

The atomic motion produced by the force $F_{\rm LZ}$ may be interpreted in the upper left energy diagram of Fig.~\ref{energyscheme} where for the case of $F_{\rm LZ}=0$ the atomic energies $E(q)$ for the $n=0,1,2$ lower bands are represented versus the quasimomentum $q$ within the Brillouin zone of width $p_{\rm B}=2p_{\rm rec}=2\pi\hbar/d_\mathrm{L}$~\cite{Cristiani02,Morsch06}. Under the action of a constant force $F_{\rm LZ}$, the quasimomentum of a condensate initially prepared at $q=0$ in the $n=0$ band scans the lower band in an oscillating motion periodically with the Bloch period $T_{\rm B}=2\hbar(F_\mathrm{LZ}d_{\rm L})^{-1}$. At the edge of the Brillouin zone, where a level splitting $\Delta E$ increasing with $V$~\cite{Holthaus00} takes place, tunneling of the condensate to the $n=1$ energy band may occur.

The Wannier--Stark Hamiltonian of Eq.~\eqref{WannierStark} can be written in dimensionless units~\cite{Holthaus00,Wimberger06}
\begin{equation}
{H_\mathrm{0}=-\frac{1}{2}\frac{\partial^2}{\partial x_\mathrm{0}^2}+\frac{V_\mathrm{0}}{16}\cos (x_\mathrm{0})+\frac{F_\mathrm{0} x_\mathrm{0}}{16\pi},}
\label{eqno5}
\end{equation}
where $ x_\mathrm{0}=2\pi x/d_{\rm L}$, and energy and time are rescaled by $H_\mathrm{0}=\tilde H/(8E_{\rm rec})$ and $t_\mathrm{0}=8tE_{\rm rec}/\hbar$ respectively. Moreover, in dimensionless units, the lattice depth is given by $V_\mathrm{0}=V/E_{\rm rec}$ and the force  by $F_\mathrm{0}=F_{\rm LZ} d_{\rm L}/E_{\rm rec}$. The translational symmetry of the given Hamiltonian, broken by the static force, is recovered using a gauge transformation. Substituting $\tilde\psi(x_\mathrm{0},t_\mathrm{0})=\exp(-iF_\mathrm{0}t_\mathrm{0}x_\mathrm{0}/16\pi)\psi(x_\mathrm{0},t_\mathrm{0})$, the Schr\"odinger equation reads $i\partial\psi/\partial t_\mathrm{0}=H(t_\mathrm{0})\psi$, with $H(t_\mathrm{0})$ the time-dependent Hamiltonian
\begin{equation}
{H(t_\mathrm{0})=\frac{1}{2}\left(\hat{p}-\frac{F_\mathrm{0}t_\mathrm{0}}{16\pi}\right)^2+\frac{V_\mathrm{0}}{16}\cos (x_\mathrm{0}),
\label{eqno6}}
\end{equation}
and the momentum operator $\hat{p}=-i\partial/\partial x_\mathrm{0}$. In the following, we analyze the Hamiltonian of Eq.~\eqref{eqno6} in the momentum basis. In order to decompose the Hilbert space into independent subspaces, we use the Bloch decomposition and for that we identify the momentum eigenstates of the free particle ($V_{\rm 0}=0=F_{\rm 0}$) for fixed quasimomentum $q$ within the Brillouin zone, i.e., $p=q+n$, $p$ and $q$ being indices in the momentum and quasimomentum representations and $n\in\mathbf Z$. To calculate the time evolution of any momentum eigenstate $|p\rangle=|q+n\rangle$, we only need the  Hamiltonian $H_{\rm q}$ acting on the subspace with a given quasi-momentum index $q$, as there is no transition between states with different $q$
\begin{equation}
  H_{\rm q} = \frac{1}{2} \left(
    \begin{array}{ccccc}
      \ddots & & & & 0 \\
        & (\tilde q-1)^2 & V_\mathrm{0}/16 & &  \\       %(q+(i-1)-F_\mathrm{0}t)^2
        & V_\mathrm{0}/16 & (\tilde q)^2 & V_\mathrm{0}/16 &  \\  %(q+i-F_\mathrm{0}t)^2
        & & V_\mathrm{0}/16 & (\tilde q+1)^2 &  \\       %(q+(i+1)-F_\mathrm{0}t)^2
        0 & & & & \ddots \\
    \end{array}
  \right)\ ,
  \label{eqno8}
\end{equation}
where $\tilde q=q-F_\mathrm{0}t_\mathrm{0}/16\pi$.

The full dynamics of the Wannier--Stark system can be locally approximated by a simple two-state model
\begin{equation}
h_{\rm q}= \frac{1}{2} \left(
    \begin{array}{cc}
         \tilde{q}^2 & V_\mathrm{0}/16 \\
          V_\mathrm{0}/16 &  (\tilde q+1)^2\\
    \end{array}
  \right)\ .
  \label{eqno9}
\end{equation}

$h_{\rm q}$ can be brought into the form of the Hamiltonian given by Eq.~\eqref{eqno1} by properly shifting the diagonal parts (e.g., shifting away the quadratic term in time $t_\mathrm{0}$). For $q=0$ we thus immediately obtain:

\begin{equation}
\frac{1}{8} \left(
    \begin{array}{cc}
         2F_\mathrm{0}E_{\rm rec}t_\mathrm{0}/\pi\hbar  & V_\mathrm{0}/4 \\
          V_\mathrm{0}/4 &   -2F_\mathrm{0}E_{\rm rec}t_\mathrm{0}/\pi\hbar \\
    \end{array}
  \right)\ .
  \label{eqno10}
\end{equation}

\indent The $\alpha$, $\Delta E$ and $\gamma$ introduced in the LZ model of Eq.~\eqref{eqno1} can be expressed in terms of our system parameters: $\alpha=2F_\mathrm{0}E^{2}_{\rm rec}/\pi\hbar=4E_{\rm rec}/(\pi T_{\rm B})$, $\Delta E=V_\mathrm{0}E_{\rm rec}/2$, and  $\gamma=32 F_\mathrm{0}/\pi V^{2}_\mathrm{0}$.
The LZ theory predicting the asymptotic behavior of the tunneling probability, can be used as a very good approximation for our system for times far enough from the avoided crossings. However there are some limiting cases, and experimental parameters, for which the simplified two-state model is not a good approximation for the Wannier--Stark system. The discrepancy is large for lattice depths larger than the energy scale $E_\mathrm{rec}$ of the system ($V_\mathrm{0} \gg 1$), where the gap between energy bands increases leading to quasi-flat bands and localized eigenstates. Therefore, several momentum eigenstates contribute with a non-negligible amount to the lowest energy eigenstate, and one would need to take into account more components in the Hamiltonian matrix.\\

\begin{figure}[htc]
 \includegraphics[width=\linewidth, angle=0]{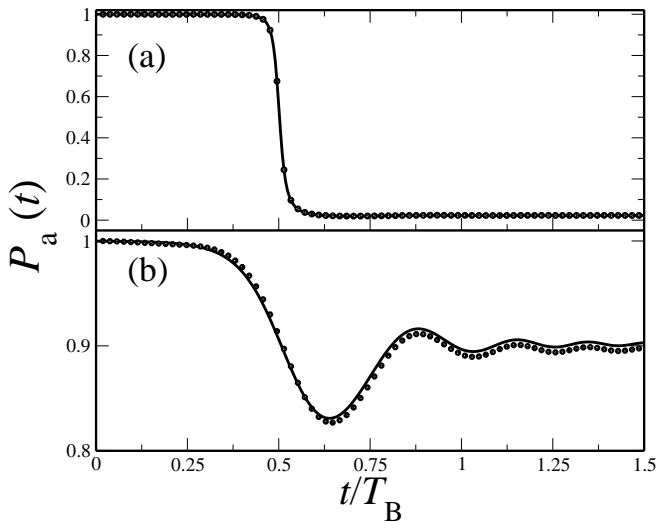}
\caption{ \small \label{fig:4} Comparison between the time evolution of the Bose--Einstein condensate survival probability, in the (a) diabatic and (b) adiabatic basis, for different initial conditions, prepared at a temporal distance $\Delta t = T_\mathrm{B}/2$ from the crossing point. The dashed line is the evolution in a possible experimental setup, i.e. the evolution following an initial preparation in the ground state of the adiabatic basis; the solid line is the evolution following an initial preparation in the ground state of the diabatic basis. Parameters are $F_{0} = 1.197$ and $V_{0} = 2.3$, corresponding to $\gamma = 2.3$, leading to a jump time in both adiabatic and diabatic bases 1.9 times the Bloch period $T_\mathrm{B}$.}
\end{figure}

\begin{figure}[htc]
\begin{center}
 \includegraphics[width=\linewidth, angle=0]{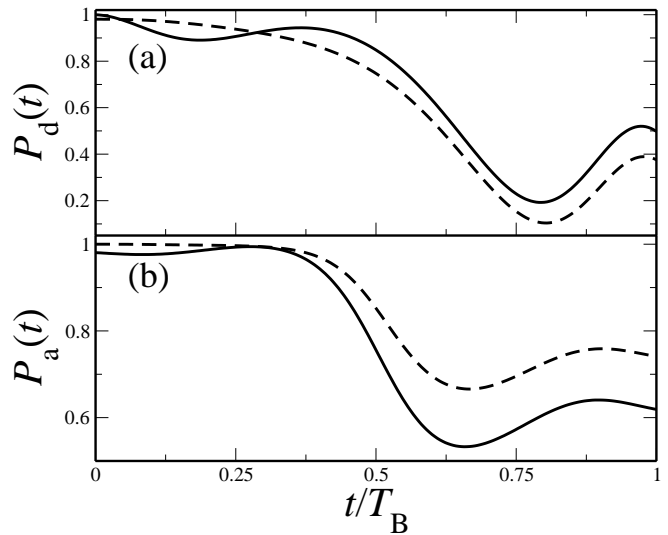}
\caption{ \small \label{fig:5}
Comparison between the time evolution of the Bose--Einstein condensate adiabatic survival probability, starting from initial ground states prepared at different time distances from the transition point. Survival probabilities measured for $F_{0}=1.197$ and $V_\mathrm{0}=0.3$ in (a), and for  $F_{0}=1.197$ and $V_{\rm 0}=3.0$ in (b). The dotted lines show the evolution obtained evolving the survival probability from the ground state in the adiabatic basis at $t=-\infty$; the solid lines illustrate the evolution obtained evolving the survival probability from the ground state which simulates a possible experimental initial state, i.e., the ground state in an adiabatic basis, at a finite time from the transition point.}
\end{center}
\end{figure}

 \subsection{Initial conditions}
Before analyzing the experiment results, we need to address an additional problem, the finite coupling duration, as in~\cite{vitanov_96}: an experiment necessarily takes a finite time for the measurement, whereas the standard LZ theory assumes that the time taken for the transition runs from $-\infty$ to $\infty$. The experimental finiteness of the sweep time $T_{\rm B}$ implies that for the initial state at a finite distance from the transition point, the diagonal and off-diagonal matrix elements in the system Hamiltonian are comparable. The experiment we are dealing with typically operates in the regime defined in~\cite{vitanov_96} as a large time, meaning that the time intervals from the turn-on and the turn-off times to the crossing are larger than the jump time.  The presence of a jump time comparable to the Bloch time may modify the temporal evolution of the survival probability  for the two mechanisms discussed in the following. Because at large $\gamma$ the ratio $\tau_{\rm d}^{jump}/T_{\rm B}$ between jump time and sweep time is given by $\pi \sqrt{F_0}/2$, large $F_0$ values may produce deviation from the ideal theory of~\cite{vitanov_99}.\\
\indent  At $t=-\infty$ the diabatic and adiabatic states coincide, and hence, the preparation of the system in its ground state is unambiguous. On the other hand, at a finite distance from the transition point, the diabatic and adiabatic states do not coincide. In a numerical approach any chosen initial state can be evolved given the proper Hamiltonian, but from an experimental point of view the system will be prepared in a well defined initial state, which depends both on the parameter values and on the preparation protocol. It is not obvious that this initial state can be chosen at will: most likely, the experimental initial state will be the one corresponding to the ground state of the complete Hamiltonian, i.e., the adiabatic lower energy state, at a time equal to the time when the sweep starts. The comparison between experiments and theory performed for different initial states should clarify this issue, because the evolution for different initial states is markedly different, both when observed in the diabatic basis and in the adiabatic basis, see Fig.~\ref{fig:4}$(a)$ and $(b)$ corresponding to  typical Bose--Einstein condensate experimental parameters. The results of Fig.~\ref{fig:4} show that, for experimentally accessible parameters, the two evolutions do not coincide in both the diabatic basis and adiabatic bases (see Fig.~\ref{fig:4}$(a)$ and $(b)$, respectively).  We have verified that the results of  Fig.~\ref{fig:4} following an initial preparation in the ground state of the diabatic basis (solid lines) coincide with the finite coupling duration predictions of ref.~\cite{vitanov_96}.\\
\indent It is not at all obvious that an initial state chosen as the adiabatic ground state at a finite time from the transition point (which is likely to be the initial experimental state) should coincide with the state obtained evolving from $t=-\infty$, projected onto the adiabatic basis. We computed the survival probability simulating different Bose--Einstein condensate initial states, see Fig.~\ref{fig:5}. For our experimental parameter set, the discrepancy is not very large, but certainly important for a precise description of the temporal evolution of the tunneling. Therefore, the approach of~\cite{vitanov_99} yields some elegant theoretical results for the LZ transition, but care is needed in comparing them with the experiment due to the presence of the additional time scale connected to the finite distance between the experimental starting point and the transition point.\\

\subsection{Numerical calculation}\label{numerics}

\begin{figure}[htc]
 \includegraphics[width=\linewidth, angle=0]{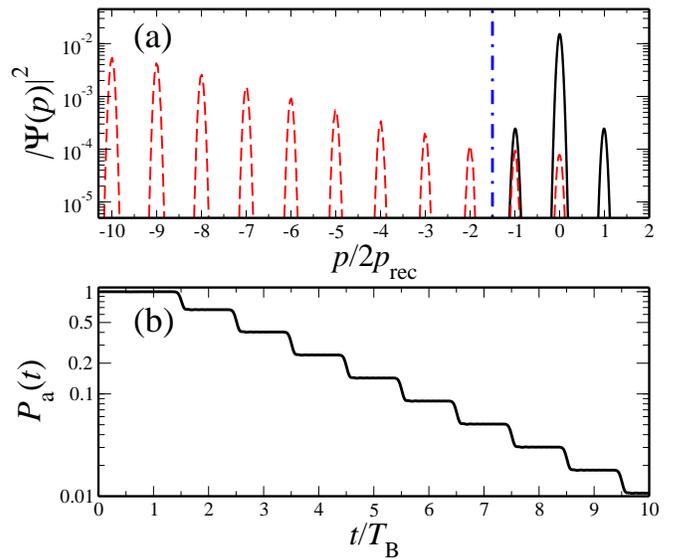}
\caption{ \small \label{fig:6} (Color online) In $(a)$ the momentum distribution at time $t=10 T_{\rm B}$ (dashed lines) starting from the initial momentum distribution (solid lines) under the action of a force directed towards negative $p$ values. The vertical dash-dotted line shows the cut-off value $p_{\rm c}=3 p_{\rm rec}$ in the definition of Eq.~\eqref{eqno11}. $(b)$ Temporal evolution of the survival probability in the adiabatic basis using mentioned definition. Simulation parameters: $V_{\mathrm 0}=2.07$, $F_{\mathrm 0}=1.197$.}
\end{figure}

In \cite{Wimberger05} some of us have introduced an easily computable quantity to determine in a good approximation the survival probability in the adiabatic basis:
\begin{equation}
P_{\rm a}(t)=\int_{-p_{\rm c}}^{\infty} dp \left|\Psi(p,t)\right|^{2},
\label{eqno11}
\end{equation}
where $\Psi(p,t)$ is the Bose--Einstein condensate wave function in momentum representation, and $p_{\rm c}\geq 3 p_{\rm rec}$ is an ad hoc cut-off. Eq.~\eqref{eqno11} can be interpreted as the projection of $\Psi(p,t)$ onto the support of the initially prepared condensate at $t=0$ (in the presence of the optical lattice but at $F_{\rm LZ}=0$), which is illustrated in Fig.~\ref{fig:6}$(a)$. Since Eq.~\eqref{eqno11} measures the decay only once the Bose--Einstein condensate wave packet $\Psi(p,t)$ has extended beyond $-p_{\rm c}$ ($=-3 p_{\rm rec}$ in Fig ~\ref{fig:6}$(a)$), we must resort to the acceleration theorem~\cite{Dahan-Peik, Holthaus00} to identify time $t$ with $t-T_{\rm B}$, i.e., we must rescale time by the traversal time of the Brillouin zone $T_{\rm B}$.\\
\indent While many previous experimental results proved in very good agreement with simulations based on Eq.~\eqref{eqno11}, c.f.~\cite{Sias07,Zenesini08,Zenesini09}, a better numerical method is needed for the new generation of experiments reported here. The dash-dotted lines in Fig.~\ref{fig:adiab_exp} were produced using $P_{\rm a}(t)$ of Eq.~\eqref{eqno11}. These simulations well reproduce the height of the steps in agreement with the LZ prediction given in Eq.~\eqref{eqno2}. They do not, however, reproduce the oscillations of the experimentally measured survival probability, due to the artificial cut-off used for evaluating $P_{\rm a}(t)$. While the sequence of steps -- corresponding to a sequence of LZ tunneling events -- is observable in Fig~\ref{fig:6}$(b)$, no oscillations are visible. To reproduce the oscillatory behavior of the experimental data in Fig.~\ref{fig:adiab_exp}, instead of Eq.~\eqref{eqno11} we determine $P_{\rm a}(t)$ in the following way: $\arrowvert\phi(n,q)\rangle$ shall denote the band solution for the ground band $n=0$ as shown in the lower left panel of Fig~\ref{fig:expscheme}. Then the adiabatic survival probability is just the projection of the condensate wave function $\Psi(p,t)$ onto $\phi(n=0,q)$ integrated over the full Brillouin zone, i.e.,
\begin{equation}
P_{\rm a}(t)=\int_{-p_{\rm rec}}^{p_{\rm rec}} dq |\langle\Psi(p=q,t)|\phi(0,q)\rangle|^{2}.
\label{eqno12}
\end{equation}

The survival probabilities $P_{\rm a}(t)$ shown in figs.~\ref{fig:4} and~\ref{fig:5} have been calculated in this way.

On the other hand, following the procedure sketched in the lower right panel of Fig~\ref{fig:expscheme}, the survival probability determined in the diabatic basis of free momentum eigenstates is given by
\begin{equation}
P_{\rm d}(t)=\int_{-p_{\rm rec}}^{p_{\rm rec}} dq |\langle\Psi(p=q,t)|p=q\rangle|^{2},
\label{eqno13}
\end{equation}
with $p=q$ within the first Brillouin zone in the notation of Section IIA. Eq.~\eqref{eqno13} is used to simulate the experimental results of Figs.~\ref{fig:diab_exp}$(a)$ and $(b)$ presented in the next section.

\section{Results}
\label{results}
\begin{figure}[htc]
 \begin{center}
 \includegraphics[width=1\linewidth,angle=0]{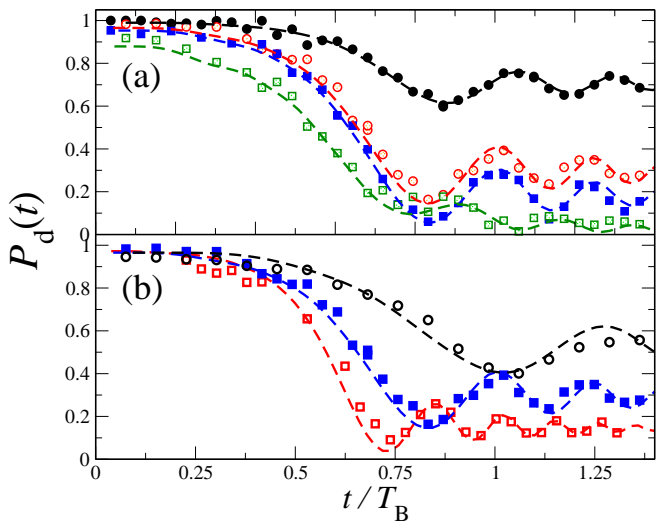}
 \caption{ \small \label{fig:diab_exp} (Color online) Time-resolved measurements of LZ tunneling in the diabatic basis. a) Fixed force $F_0=1.197$ with different lattice depths $V_0=1$ (filled circles), $1.8$ (open circles), $2.3$ (filled squares) and $4$ (open squares). b) Fixed lattice depth $V_0=1.8$ with different forces: $F_0=2.394$ (open circles), $1.197$ (filled squares) and $0.599$ (open squares). The dashed lines are the results of numerical simulations based on Eq.~\eqref{eqno13} which nicely reproduce the experimental data.}
 \end{center}
 \end{figure}

\indent In our experiments we realized the Wannier--Stark Hamiltonian of Eq. (\ref{WannierStark}) with Bose-condensed rubidium atoms inside an optical lattice \cite{Morsch01,Cristiani02,JonaLasinio03,Sias07,Zenesini08,Zenesini09}. Initially, we created Bose--Einstein condensates of $5\times 10^4$ $^{87}$Rb atoms inside an optical dipole trap (mean trap frequency around $80\,\mathrm{Hz}$). A one-dimensional optical lattice created by two counter-propagating, linearly
polarized gaussian beams was then superposed on the Bose--Einstein condensate by ramping up the power in the lattice beams in $100\,\mathrm{ms}$. The wavelength of the lattice beams was $\lambda=842\,\mathrm{nm}$, leading to a sinusoidal potential with lattice constant $d_{\rm L}=\lambda/2=421\,\mathrm{nm}$. A small frequency offset $\Delta \nu(t)$ between the two beams could be introduced through the acousto-optic modulators in the setup, which allowed us to accelerate the lattice in a controlled fashion and hence, in the rest-frame of the lattice, to subject the atoms to a force $F_{\rm LZ}=Ma_\mathrm{LZ}$ with $a_\mathrm{LZ}=d_\mathrm{L}\frac{d\Delta \nu(t)}{dt}.$\\
\indent In several previous experiments~\cite{Cristiani02,JonaLasinio03,Sias07,Zenesini08} we had already measured the LZ tunneling probability by first loading the Bose--Einstein condensate into a lattice, then accelerating the lattice for one Bloch period (i.e., across the zone edge and then to the center of the second Brillouin zone) and subsequently measuring the number of atoms left in the fundamental band. This was done by accelerating the lattice further with a smaller value of $a_\mathrm{sep}$ and a larger lattice depth $V_\mathrm{sep}$ chosen such as to ensure that atoms in the fundamental band did not undergo LZ tunneling at subsequent crossing of the zone edge and that atoms in higher bands tunneled with almost 100\% probability. In that way it was possible to separate atoms in the fundamental band in momentum space so that after a time-of-flight they could be easily measured.\\
\indent The time-resolved measurements we are interested in for the purposes of the present paper initially followed the same procedure. Rather than accelerating the lattice for a full Bloch period, however, we had to interrupt the LZ tunneling event at some time $t\neq nT_\mathrm{B}$ in general. The exact protocol then depended on whether we wanted to measure in the adiabatic or in the diabatic basis.\\
\indent For measurements in the {\em adiabatic} basis, we proceeded as follows, see Fig.~\ref{fig:expscheme}. After loading the Bose--Einstein condensate into the optical lattice, the lattice was accelerated with acceleration $a_\mathrm{LZ}$ for a time $t_\mathrm{LZ}$. The lattice thus acquired a final velocity $v=a_\mathrm{LZ} t_\mathrm{LZ}$. At time $t=t_{\mathrm LZ}$ the acceleration was abruptly reduced to a smaller value $a_\mathrm{sep}$ and the lattice depth was increased to $V_\mathrm{sep}$ in a time $t_\mathrm{ramp}\ll T_\mathrm{B}$. These values were chosen in such a way that at time $t=t_\mathrm{LZ}$ the probability for LZ tunneling from the lowest to the first excited energy band dropped from between $\approx 0.1-0.9$ (depending on the initial parameters chosen) to less than $\approx 0.01$, while the tunneling probability from the first excited to the second excited band remained high at about $0.95$. This meant that at $t=t_\mathrm{LZ}$ the tunneling process was effectively interrupted and for $t>t_\mathrm{LZ}$ the measured survival probability $P(t)=N_0/N_\mathrm{tot}$ (calculated from the number of atoms $N_0$ in the lowest band and the total number of atoms in the condensate $N_\mathrm{tot}$) reflected the instantaneous value $P(t=t_\mathrm{LZ})$.\\
\indent The lattice was then further accelerated for a time $t_\mathrm{sep}$ such that $a_\mathrm{sep}t_\mathrm{sep}\approx 2m p_\mathrm{rec}/M$ (where typically $m=2$ or $3$). In this way, atoms in the lowest band were accelerated to a final velocity $v\approx 2m p_\mathrm{rec}/M$, while atoms that had undergone tunneling to the first excited band before $t=t_\mathrm{LZ}$ underwent further tunneling to higher bands with a probability $>0.95$ and were, therefore, no longer accelerated. At time $t_\mathrm{sep}$ the lattice and dipole trap beams were suddenly switched off and the expanded atomic cloud was imaged after
$23\,\mathrm{ms}$. In these time-of-flight images the two velocity classes $0$ and $2m p_\mathrm{rec}/M$ were well separated and the atom numbers $N_0$ and $N_\mathrm{tot}$ could be measured directly. Since the populations were effectively "frozen" inside the energy bands of the lattice, which represent the adiabatic eigenstates of the total Hamiltonian of the system, this experiment measured the time dependence of the LZ survival probability $P_{\rm a}$ in the {\it adiabatic} basis, c.f. Eq.~\eqref{eqno12} above.\\
\indent The results of our measurements in the adiabatic basis are summarized in Fig. \ref{fig:adiab_exp}. The step-like behavior of the survival probability around $t=0.5T_\mathrm{B}$ is clearly visible, as well as the finite width of the step, which demonstrates that our experimental protocol does, indeed, allow us to access the dynamics of the LZ transition and the jump time associated with that transition. Also shown in the figure are the results of numerical simulations using the cut-off and the adiabatic survival methods described above in Section IIIC. As expected, both methods reproduce the step with a finite width and the steady-state value of the survival probability for long times. The slight oscillations of the survival probability for $t>0.5T_\mathrm{B}$, however, are only visible in the results computed according to Eq.~\eqref{eqno12} above. In fact, the amplitude of these oscillations is larger in the numerical simulations than in our experimental data. This might indicate that our protocol for freezing the instantaneous populations in the ground and excited bands is not perfect. Indeed, we found that a delicate balance between the accelerations and lattice depths for the separation phase was necessary in order to ensure that the populations after the separation phase faithfully reproduced those at $t=t_\mathrm{LZ}$, which was tested by choosing two extreme values for $a_\mathrm{LZ}$ which gave theoretical survival probabilities of approximately $0$ and $1$, respectively, and then verifying that these values were measured in the experiment. In practice, the parameters for the separation phase were optimized in this way for one set of the LZ parameters and then kept constant as $V$ was varied in Fig. \ref{fig:adiab_exp}.\\
\indent For measurements in the {\em diabatic} basis, the experimental protocol was even simpler, see Fig. \ref{fig:expscheme}. As in the adiabatic case, after the initial loading phase the lattice was accelerated with acceleration $a_\mathrm{LZ}$ for a time $t_\mathrm{LZ}$. At that point the atomic sample was projected onto the free-particle diabatic basis by instantaneously (within less than $1\,\mathrm{\mu s}$) switching off the optical lattice. After a time-of-flight the number of atoms in the $v=0$ and $v=2p_\mathrm{rec}/M$ momentum classes are measured and from these the survival probability (corresponding to the atoms remaining in the $v=0$ velocity class relative to the total atom number) is calculated. Fig. \ref{fig:diab_exp} shows the results of such measurements, together with numerical simulations based on Eq.~\eqref{eqno13}. As in the adiabatic case, a step of the survival probability around $t=0.5\,T_\mathrm{B}$ is clearly seen, as well as strong oscillations for $t>0.5\,T_\mathrm{B}$. These oscillations are much stronger and visible for a wider range of parameters in the diabatic basis than in the adiabatic basis (see the results for $V_0=2.3$ in Fig. \ref{fig:adiab_exp}, which is confirmed by our numerical simulations).\\

 \section{Conclusions}

\indent Ultracold atoms in optical lattices provide an ideal model system for time-resolved studies of LZ tunneling. The complete control over the parameters of the lattice makes it possible to measure the tunneling dynamics in the adiabatic and diabatic bases by using different measurement methods. Our results confirm the existence of a finite temporal width for the transition in both bases and of strong oscillations of the survival probability in the diabatic basis. Both of these features are backed up by numerical simulations taking into account details of the experimental protocol.\\
\indent Our findings pave the way towards more quantitative studies of the tunneling time for LZ transitions, which are of current interest in the context of optimal quantum control and the quantum speed limit \cite{QSL}. Also, it should be possible to measure the tunneling dynamics in arbitrary bases by inducing a rotation of the $2\times 2$ LZ-matrix through variations in the lattice depth during the transition. With an appropriate choice of this variation one could then, for instance, realize the super-adiabatic basis proposed by M. Berry \cite{Berry_90}.\\

{\bf ACKNOWLEDGMENTS}
\indent We gratefully acknowledge funding by the EU project "NAMEQUAM", the CNISM "Progetto Innesco 2007" and the Excellence Initiative by the German Research Foundation (DFG) through the Heidelberg Graduate School of Fundamental Physics (grant number GSC 129/1) and the Global Networks Mobility Measures. G. Tayebirad thanks the Landesgraduiertenf\"{o}rderung Baden-W\"{u}rttemberg for support.

\bibliographystyle{apsrmp}

\end{document}